\documentclass[12pt]{JHEP3}
\pdfoutput=1
\usepackage{amsmath}

\def\eqref#1{(\ref{#1})}

\def\pa{\partial}

\def\be{\begin{equation}}
\def\ee{\end{equation}}
\def\bea{\begin{eqnarray}}

\def\eea{\end{eqnarray}}
\def\nn{\nonumber}

\def\l{\lambda}
\def\lab{\label}
\def\f{\phi}

\def\e{\epsilon}

\def\le{\left}
\def\ri{\right}

\def\t{\tau}
\def\k{\kappa}

\def\cO{{\cal O}}
\def\m{\mu}
\def\n{\nu}

\def\6{\partial}

\def\a{\alpha}
\def\b{\beta}
\def\S{\Sigma}
\def\lab{\label}

\title{Late time behavior of non-conformal plasmas }

\author{Umut G\"ursoy$^{a}$, Matti J\"arvinen$^{b,c}$, Giuseppe~Policastro$^b$
~\\ \\
$^a$ Institute for Theoretical Physics, Utrecht University \\
Leuvenlaan 4, 3584 CE Utrecht, The Netherlands
~\\ \\
$^b$ Laboratoire de Physique Th\'eorique, Ecole Normale Sup\'erieure \\
24 rue Lhomond, 75231, Paris Cedex 05, France (UMR du CNRS 8549)
~\\ \\
$^c$ Institut de Physique Th\'eorique Philippe Meyer\\
24 rue Lhomond, 75231, Paris Cedex 05, France
~\\ \\
}

\abstract{We determine analytically the dependence of the approach to thermal equilibrium of strongly coupled plasmas on the breaking of scale invariance. The theories we consider are the holographic duals to Einstein gravity coupled to a scalar with an exponential potential. The coefficient in the exponent, $X$, is the parameter that controls the deviation from the conformally invariant case. For these models we obtain analytic solutions for the plasma expansion in the late-time limit, under the assumption of boost-invariance, and we determine the scaling behaviour of the energy density, pressure, and temperature as a function of time. We find that the temperature decays as a function of proper time as $T\sim \tau^{-s/4}$ with $s$ determined in terms of the non-conformality parameter $X$ as $s=4(1-4X^2)/3$. This agrees with the result of Janik and Peschanski, $s=4/3$, for the conformal plasmas and generalizes it to non-conformal plasmas with $X\neq 0$. We also consider more realistic potentials where the exponential is supplemented by power-law terms. Even though in this case we cannot have exact solutions, we are able under certain assumptions to determine the scaling of the energy, that receives logarithmic corrections.}

\preprint{...}
\keywords{\vspace{-1cm} AdS/CFT, Quark-Gluon Plasma, thermalization}

\begin{document}

%\body

%\version\versionno
%\tableofcontents

\section{Introduction}

The study of strongly coupled systems by means of the holographic duality is by now a well-established field of research, which has produced a number of insights into the dynamics in a wide range of situations in which conventional techniques can not readily be applied for one reason or another. In particular, the duality has made the regime of  dynamics far from equilibrium much more accessible. The first studies were mostly concerned with linear response, i.e. the regime of  small fluctuations away from equilibrium, but in recent years the number of investigations into genuinely non-equilibrium phenomena has been growing steadily. 

In the linear response regime, the holographic prescription boils down to solving the linearized equations for perturbations of the metric or other fields around a given background. The full solution of the problem is then equivalent to finding all the normal or quasi-normal modes for the relevant fluctuations. 
Even at this level, analytic solutions can often be obtained only in the simplest of cases and typically one has to resort to numerics. Far from the linearized regime one has to find solutions to  full Einstein equations and the known analytic solutions are even scarcer. It was therefore a remarkable achievement when in \cite{Janik:2005zt} Janik and Peschanski found the dual solution to the flow of an expanding plasma. They considered the plasma of ${\cal N}=4$ SYM, which is the prime and most studied example of holographic duality.
In more detail, the situation considered in \cite{Janik:2005zt} is that of a boost-invariant flow. The assumption of boost-invariance was introduced by Bjorken \cite{Bjorken:1982qr} and is considered to be a good approximation to the behavior of the fluid created in the heavy-ion collisions at least in the central rapidity region 
(see \cite{Romatschke:2009im} for a more extended discussion of the validity of this assumption).  
Under the assumption of boost invariance and conformal invariance, 
the stress-energy tensor is completely determined in terms of the energy density that is a function of only one variable, the proper time. From a dual perspective, \cite{Janik:2005zt} showed that the Einstein equations in the bulk admit an expansion in terms of inverse power of time, and the leading late-time solution is found by solving a  set of \emph{ordinary} non-linear differential equations, for which they found an analytic solution. The gravity dual then predicts a behavior of the energy density that is compatible with the assumption of a perfect fluid. In a sense this was an expected result since the hydrodynamic approximation had already been amply tested in the linearized regime, but it was still a non-trivial extension of the duality to a fully dynamical situation. 

The motivation for the work of \cite{Janik:2005zt} was of course, as we alluded to, the application of the duality to the study of the quark-gluon plasma produced in the heavy-ion collisions. It is well-understood that some features of the hydrodynamical evolution are universal in the holographic setup. The value of the shear viscosity is independent of the model considered\cite{Policastro:2001yc, Kovtun:2004de, Buchel:2003tz}, as long as there are no higher derivative terms in the action\footnote{See \cite{Cremonini:2011iq} for a recent review on the shear viscosity in holographic models with higher derivative corrections.}. For some purposes however it is important to keep track of the breaking of conformal invariance that occurs in the real-world QCD plasma. 

The purpose of this note is to take a step in the direction of understanding the influence of the absence of conformal symmetry on the thermalization of the system.  In linearized hydrodynamics the absence of conformal invariance manifests itself in the presence of a \emph{bulk viscosity}, and its value can be easily determined in a large class of models. Even though the bulk viscosity grows in the vicinity of the deconfinement phase transition, it is not clear to what extent it will influence the evolution. We will be interested here in a different effect, that can be directly attributed to the trace anomaly, and so it is already present at the level of ideal fluid. We will see in fact that the presence of the trace anomaly modifies the leading late-time behavior of the decay of the energy density.

We consider a class of models with Einstein gravity coupled to a scalar field with a potential. Ideally we would like to take a potential adapted to model the features of QCD, namely confinement and asymptotic freedom, as in the Improved Holographic QCD program \cite{ihqcd1,ihqcd2}. However for a realistic choice of potential one cannot obtain analytic expressions even for the static black hole, let alone more complicated dynamical solutions. We choose instead to consider a toy model with a simple form for the potential, namely a single exponential $e^{\alpha \phi}$, where $\phi$ is the scalar field (it can be seen as a dilaton). The coefficient $\alpha$ parametrizes the breaking of scale invariance. Even though the single exponential potential is not as realistic as the potentials employed in improved Holographic QCD, it captures the large $\phi$ behavior of those potentials, possibly with subleading power law corrections in $\phi$.  Therefore it indeed corresponds to the IR limit of the improved holographic QCD potentials, hence is useful for the late time behaviors of such realistic models. The disadvantage is that such a model does not admit a stable vacuum, so the dual theory is not well-defined in the UV and it would require a UV completion; this is not a terrible drawback since we are interested in infrared properties of the system. The value of $\alpha$ determines also the presence or absence of confinement in the vacuum of the field theory. As it turns out, we are able to study only the cases that correspond to a non-confining vacuum. 

For this class of models, analytic black hole solutions are known \cite{ChamblinReall}. By adapting the method of \cite{Janik:2005zt} we can find the corresponding late-time evolving solutions, for a boost-invariant flow, again analytically. Therefore we can find the exact correction to the decay exponent of the energy density and temperature and the precise dependence on the scale-symmetry breaking. Explicitly,
we find that the temperature decays at late times as $T \sim \tau^{-s/4}$, with $s$ is given in terms of the coefficient $\alpha$ of the dilaton potential as
\be
 s = \frac{4}{3}\left[1 -\left(\frac{3\alpha}{4}\right)^2 \right] \equiv \frac{4}{3}\le(1-4X^2\ri)\,,
\ee
where we also defined the parameter $X=-3\alpha/8$ which will be used below.

We find that the decay becomes slower compared to the conformal case $s=4/3$, and the exponent approaches zero at the critical case that corresponds to a confining model. Unfortunately, as already explained, we are not able to cross over to the confining regime because our solutions become unphysical. We can say something more about the region close to the critical point by modifying the potential to include subleading powerlike terms, $e^{\alpha \phi} \phi^P$. In this case we do not have full analytic solutions, but an asymptotic solution in $1/\phi$ is sufficient to determine logarithmic corrections to the power-law decay of the energy density. 

There are a few recent papers that came out while this work was in preparation, that study  the effect of absence of scale invariance on thermalization but from different point of views than ours. We mention their results for the reader's orientation.

In \cite{Ishii:2015gia} they consider a similar class of gravity-scalar models subject to a quench induced by  specifying time dependent boundary conditions on the scalar.  The main result is that there is a dominant thermalization timescale determined by the imaginary part of the lowest quasi-normal mode of the black brane to which the system relaxes. 

In \cite{Buchel:2015ofa} they study the thermalization time by means of quasinormal modes of the transverse traceless fluctuations of the stress-energy tensor; they find a very mild dependence on the breaking the scale invariance over a large range of the parameter $\delta = \frac{1}{3}-c_s^2$.

In \cite{Janik:2015waa} they consider  holographic models with an equation of state inspired by lattice QCD, and study the behavior of the lowest non-hydrodynamical quasinormal mode. Again they find a moderate dependence of the damping of the mode on the conformal breaking, by a factor of about two between the extreme cases. 

The outline of the paper is as follows: in section \ref{Bjorkflow} we briefly recall the properties of the boost-invariant flow and its holographic description by  \cite{Janik:2005zt} for the conformal case. In section \ref{CR} we describe the (Chamblin-Reall) black-hole solutions for the model with a single exponential potential, and the corresponding late-time dynamical solutions. We determine the corresponding field theory stress-energy tensor by using dimensional reduction and the holographic renormalization. In section \ref{CRpow} we consider the case of the exponential potential modified by a power. In section \ref{Conc} we summarize our results and point out some directions for further work. The appendices contain more details on the gravity solutions and the thermodynamics of the system. 

\section{Bjorken flow in a CFT and its gravity dual}\label{Bjorkflow}

Let us review the picture of the boost-invariant flow advocated by Bjorken \cite{Bjorken:1982qr}. It is convenient 
to introduce the pseudo-rapidity and the proper-time as 
\be\lab{coors} 
t = \tau \cosh(y)\, \qquad x^1 = \tau \sinh(y)\, .
\ee
In these coordinates, the boost-invariance reduces to independence on the coordinate $y$. 
The metric in this coordinate system is 
\be\lab{metf} 
ds^2 = -d\tau^2 + \tau^2 dy^2 + dx_\perp^2 \, .
\ee
Under the assumed symmetries (boost-invariance and translational invariance in the transverse plane), 
the conservation of energy-momentum tensor gives the equation 
\be\lab{NS} 
\tau \frac{d}{d\tau} T_{\tau\tau}  + T_{\tau\tau} + \frac{1}{\tau^2} T_{yy} = 
0\, .
\ee 
The trace of the energy-momentum tensor is
\be\lab{trace2} 
T^\m_\m = -T_{\tau\tau} + \frac{1}{\tau^2}T_{yy} + 2 T_{xx} \, .
\ee
In a given theory, the equation of state will give the trace as a function of the temperature. In a CFT the trace vanishes identically, but anticipating the following sections, we will consider theories for which this function is a power: $T_\mu^\mu \propto T^\xi$. 
We assume that we are working in the adiabatic approximation so that we can use the same 
equation of state as in the equilibrium case,  with the temperature $T$ becoming time-dependent $T(\tau)$. 
The trace equation then reads 
\be\lab{eq1} 
 -T_{\tau\tau} + \frac{1}{\tau^2}T_{yy} + 2 T_{xx}  = - c\, T^\xi \, . 
\ee

Defining $T_{\tau\tau} = \e(\tau)$ we find 
\be\lab{emtCR}
T_{\m\n} = \textrm{diag}\, \le( \e(\tau),\, -\tau^3 \6_\tau \e - \tau^2\e,\, \e 
+ \frac{\tau}{2} \6_\tau \e - \frac{c}{2} T^\xi ,\e + \frac{\tau}{2} \6_\tau 
\e - \frac{c}{2} T^\xi \ri)\, .
 \ee
 Now, if we further impose the {\em perfect fluid} form  
 \be\lab{pf} 
 T^{\m\n} = (\e + p) u^\mu u^\n + p \eta^{\m\n} \, ,
 \ee
we have another condition on the components of the stress-energy tensor: 
$T_{xx} = \tau^{-2} T_{yy}$. 
Using (\ref{emtCR}) in this equation we can solve for the energy $\e(\tau)$ as 
\be\lab{sole}
\e(\tau) = \e_0 \tau^{-\frac43} +\frac{c}{2} \tau^{-\frac43} \int_{\tau}^\infty 
d\tilde{\tau} \tilde{\tau}^{\frac13} T(\tilde{\tau})^\xi\, .
\ee
The integral can be performed if we assume a power-law behavior for $T(\tau)$:  
\be\lab{solepar}
T = T_0 \tau^{-\a} \,\,  \Rightarrow \,\, \e(\tau) = \e_0 \tau^{-\frac43} +\frac{c\, T_0^\xi}{4-3\a \xi} 
\tau^{-\a\xi}\, .
\ee
The conformal case is obtained by setting $c=0$ and results in an energy decay with time with exponent 4/3. Since 
scale invariance implies that $\epsilon\sim T^4$, it follows that the temperature decreases as $T \sim \tau^{-1/3}$. 
In the non-conformal case, we notice that the contribution from the trace anomaly will dominate the late-time behavior if 
\be\lab{req} 
\alpha\xi < \frac43\, .
\ee

We now outline the idea of \cite{Janik:2005zt} that we will follow closely in this paper. They start by considering the 
most general Ansatz for a bulk metric in $AdS_5$ consistent with the symmetries of the Bjorken flow; this has the form 
\be\label{JPansatz}
ds^2 = \frac{1}{z^2} \left( dz^2 - e^{a(z,\tau)} d\tau^2 + \tau^2 e^{b(z,\tau)} + e^{c(z,\tau)} dx_\perp^2 \right) \,. 
\ee
They found that one can consistently assume a scaling behavior in terms of a coordinate $$v = \frac{z}{\tau^{s/4}}$$ 
with an unspecified constant $s$. The metric functions $a,b,c$ are then functions of $v$ up to corrections suppressed by powers of $\tau$, so this scaling Ansatz describes the late time behavior of the system. The Einstein equations reduce to a set of coupled non-linear differential equations that can be solved by the following change of variables: 
\begin{eqnarray} \label{CRchange}
a(v) &=& A(v) - 2 m(v) \,,\nonumber \\
b(v) &=& A(v) + (2s-2) m(v) \,, \\
c(v) &=& A(v) + (2-s) m(v) \,.\nonumber
\end{eqnarray}
The solution is given by 
\begin{eqnarray} 
A(v) &=&  \frac{1}{2} \log (1- \Delta(s)^2 v^8) \,, \nonumber \\
m(v) &=& \frac{1}{4 \Delta(s)} \log \frac{1+\Delta(s) v^4}{1-\Delta(s) v^4}
\end{eqnarray}
where $\Delta(s) = \sqrt{\frac{3s^2-8s+8}{24}}$. Such solutions correspond to a boundary energy density behaving as $\epsilon \sim \tau^{-s}$, and there is a solution for generic $s$, however the form of the solution shows a potential singularity at $v^4 = 1/\Delta(s)$. The analysis of the curvature invariants shows that there is indeed a singularity except for a specific value of $s$, namely $s=4/3$. This, as we have seen, is the behavior expected for a perfect conformal fluid, therefore this analysis showed, in a dynamical setup, that the fluid dual to $AdS$ gravity is a perfect fluid to leading order in the late-time expansion (which corresponds to the hydrodynamical derivative expansion). 
The subleading corrections in $1/\tau$ contain informations about the deviation from perfect fluid, in particular the viscosity coefficients \cite{Janik:2006ft}, which we will not consider here.

\section{Chamblin-Reall plasma} \label{CR}
\subsection{Black brane solution} 

We consider the Einstein-dilaton theory in 5 dimensions given by the following action 
\begin{equation}\label{action}
 {\cal A} = \frac{1}{16\pi G_5} \int d^{5}x \sqrt{-g}\le( R - \frac43 (\6\phi)^2 
+ V(\phi)
  \ri)+\,\,\, G.H.
\end{equation}
where $G.H.$ stands for the Gibbons-Hawking term, and with a single exponential potential for the dilaton
\begin{equation}\label{exppot}
    V  = V_0 (1-X^2) e^{-\frac83 X\, \phi}\,.
\end{equation}
Here $X$ and $V_0$ %and $\xi$ 
are constants.  The parameter $X$ determines the running of the dilaton and hence the breaking of conformal invariance. Without loss of generality we take $X<0$. In \cite{ChamblinReall} Chamblin and Reall found an analytic black brane solution to this system with a non-trivial profile for the dilaton field. We refer to the dual finite-temperature state as a \emph{CR plasma}. The analytic {\em black-brane} and a {\em thermal gas} (no-horizon) solution of 
this action can be expressed in terms of the following metric functions
\be\lab{dw} 
ds^2 = e^{2A(u)} \le(-f(u) dt^2 + \delta_{ij} dx^{i} dx^{j} \ri) + 
\frac{du^2}{f(u)}, \qquad \phi=\phi(u)\, . 
\ee
The dilaton has the same form both in the black-hole and the thermal gas solution: 
\begin{equation}\label{dilsoll}
    \l \equiv e^{\f} = \le(C_1 - 4X^2\frac{u}{\ell}\ri)^{\frac{3}{4X}}\,,
\end{equation}
where $\ell = \sqrt{12/V_0}$, and the scale factor is
\begin{equation}\label{Asol}
    e^A = e^{A_0} \l^{\frac{1}{3X}}\, .
\end{equation}
For the thermal gas the blackening factor $f(u)=1$, whereas for the black-hole solution 
\begin{equation}\label{fsol}
 f(u) = e^g = 1- C_2\l^{-\frac{4(1-X^2)}{3X}}\, .
\end{equation}
The boundary is located at $u = -\infty$. In order for $f$ to be a monotonically 
decreasing function one must require 
\be\lab{Xcon}
-1 < X < 0 \,. %\frac12\, .
\ee
Here $C_1$, $C_2$ and $A_0$ are integration constants: $C_1$ is the location of 
the singularity, $C_2$ determines the location of the horizon. In terms of dual 
theory one can think of $A_0$  determining the size of the dual plasma (or the 
string tension), $C_1$ determining some conformality breaking scale 
$\Lambda_{QCD}$ and $C_2$ the temperature $T$ of the plasma.   For the thermal 
gas we set $C_2=0$.  

We find $f\to 1$ on the boundary, ($\l\to 0$) as long as $-1<X<0$. 
There is an event horizon located at (using (\ref{dilsoll})),
\begin{equation}\label{ehor}
    \l_h = C_2^{\frac{3X}{4(1-X^2)}}\qquad i.e. \qquad \frac{u_h}{\ell} =
    \frac{C_1}{4X^2}- \frac{C_2^{\frac{X^2}{1-X^2}}}{4X^2}\, .
\end{equation}
The curvature singularity is located at $\l=\infty$ i.e.,
\begin{equation}\label{cursin}
    \frac{u_0}{\ell}  = \frac{C_1}{4X^2}\,.
\end{equation}
We note that when $C_2 \neq 0$ then $u_h<u_0$ and indeed there is a well-behaved 
black-hole
solution to the system. The metric of the black-hole is given by,
\bea
    ds^2 &=& e^{2A_0}\le(C_1-4X^2\frac{u}{\ell}\ri)^{\frac{1}{2X^2}}
    \le\{dx_idx^i -    
\le(1-C_2\le(C_1-4X^2\frac{u}{\ell}\ri)^{-\frac{1-X^2}{X^2}}\ri)dt^2\ri\}\nonumber\\
   {}&& + 
\le(1-C_2\le(C_1-4X^2\frac{u}{\ell}\ri)^{-\frac{1-X^2}{X^2}}\ri)^{-1}du^2\label{BHmet}.
\eea
The temperature of the black-hole is determined by requiring
regularity of the Euclidean continuation at $u_h$:
\begin{equation}\label{tempgen}
    \b=\frac{1}{T} = \frac{4\pi}{|f'(u_h)|e^{A(u_h)}}\,.
\end{equation}
One finds,
\begin{equation}\label{temp}
    \b = \pi \ell
    \frac{e^{-A_0}C_2^{-\frac{\frac14-X^2}{1-X^2}}}{1-X^2}\,.
\end{equation}
From this formula one sees that $X = -1/2$ is a threshold value. When $X$ goes below this value the temperature increases as the horizon size decreases, so one is on the small black hole branch, which is thermodynamically unstable. 
On the threshold value for $X$, i.e. $X = -1/2$, interestingly the
temperature is completely fixed by the integration constant $A_0$:
\begin{equation}\label{tempsp}
    \b = \frac1T = \frac{4\pi \ell}{3 e^{A_0}}\,.
\end{equation}
Otherwise the temperature is determined by the combination of $A_0$
and $C_2$, namely the string tension and the location of the event
horizon.

The thermal gas solution is found by setting $C_2=0$, hence $f=1$. The dilaton 
is given again by
(\ref{dilsoll}) and the metric is,
\begin{equation}\label{TGmet}
    ds^2 = e^{2A_0}\le(C_1-4X^2\frac{u}{\ell}\ri)^{\frac{1}{2X^2}}
    \le\{dx_idx^i + dt^2\ri\} + du^2\,.
\end{equation}
Here we required the same integration constant for $A$ as the
black-hole solution (\ref{BHmet}). This is because they should have
the same asymptotics at the boundary. Euclidean time is compactified
with circumference, $\bar{\b}$. We note that there is a curvature
singularity at $u_0$ that is given by (\ref{cursin}). It is the same
locus as the curvature singularity of the black-hole solution -- that
is cloaked behind the event horizon -- resides.

When $-1/2<X<0$ we can also compute the solution in the conformal coordinate system: 
\be\lab{conf} 
ds^2 = e^{2A(r)} \left(-f(r) dt^2 + \delta_{ij} dx^{i} dx^{j}  + 
\frac{dr^2}{f(r)}\right), \qquad \phi=\phi(r)\, .
\ee
This is easily obtained from the solution above by the change of variables $du = 
e^{A} dr$: 
\bea\label{CRconf1}
    \l \equiv e^{\f} &=& \le(\frac{r}{\ell'}\ri)^{-\frac{3 X}{1-4X^2}}, \qquad 
e^{A(r)} =e^{A_0} \l(r)^\frac{1}{3X}, \\
    f(r) &=& 1- \le(\frac{r}{r_h}\ri)^{\frac{4(1-X^2)}{1-4X^2}}\qquad \ell' = 
\frac{\ell}{e^{A_0}  (1-4X^2)}\, .
    \eea   
Here the boundary is located at $r=0$ and the location of the horizon is $r=r_h$. 
The temperature of the black-brane solution is given in terms of $r_h$ as 
\be\lab{Trh} 
T = \frac{1}{\pi r_h} \frac{1-X^2}{1-4X^2}\, .
\ee

\subsection{Thermodynamics of the CR plasma} \label{sec:CRthermo}

The entropy (density) of the black-brane is determined from the area of the 
horizon as 
\be\lab{ent} 
S = c_s \le(T\ell\ri)^{\frac{3}{1-4X^2}}, \qquad  c_s = \frac{e^{3A_0}}{4G_5} 
\le(\frac{e^{A_0}(1-X^2)}{\pi}\ri)^{\frac{3}{4X^2-1}}\, .  
\ee
The free energy is obtained from the first law as $F = -\int S dT$. One finds 
\be\lab{free} 
F = -c_f \le(T\ell\ri)^{\frac{4(1-X^2)}{1-4X^2}}, \qquad  c_f = \frac{c_s}{\ell} 
\frac{1-4X^2}{4(1-X^2)}\, .  
\ee
The energy is given by $\epsilon = F + TS$ as 
\be\lab{free2} 
\epsilon = c_e\le(T\ell\ri)^{\frac{4(1-X^2)}{1-4X^2}}, \qquad  c_e = \frac{c_s}{\ell}  
\frac{3}{4(1-X^2)}\, .  
\ee
The trace of the energy-momentum tensor is given by 
\be\lab{trace1} 
-T^\m_\m = \epsilon + 3F  = \frac{3c_s}{\ell} 
\frac{X^2}{1-X^2}\le(T\ell\ri)^{\frac{4(1-X^2)}{1-4X^2}} \, .  
\ee
Comparing with (\ref{eq1}) we find that in this class of models 
\be\lab{xi} 
\xi =  \frac{4(1-X^2)}{1-4X^2}\, .
\ee

Alternatively we can obtain the free energy from the action (\ref{action}) 
evaluated on-shell.  This method yields the same results as above and the details are presented in Appendix~\ref{App-FE}. In particular one obtains the following difference for the on-shell actions of the black brane and the thermal gas solutions: 
 \begin{equation}\label{sdif1}
    S_{BH} - S_{TG} = - M^3 V_3\le(\frac{\b}{\ell}\ri) e^{4A_0} 
C_2\le(1 -4 X^2\ri)\,.
\end{equation}
As (\ref{sdif1}) is negative (positive) for $-1/2<X<0$ (for $-1<X<-1/2$), the BH (TG) solution minimizes the action, hence it is the dominant solution.  Notice that {\em there is no finite temperature phase transition in this geometry}.

\subsection{Bulk viscosity of the plasma} 

One important difference between the non-conformal plasmas that we consider in this paper and the conformal ones 
is that the dissipation in these systems is characterized both by the shear viscosity and the bulk viscosity. The latter vanishes for conformal plasmas by scale invariance. The bulk viscosity of generic black brane solutions was first obtained in \cite{TarrioMas}. Adapting their formula to our normalization of the $\phi$ kinetic term in  (\ref{action}) we obtain for the following bulk-viscosity to entropy ratio for theories with potential (\ref{exppot}): 
\be\lab{bv1}
\frac{\zeta}{S}  = \frac{1}{4\pi}\sqrt{\frac83}\, X\, .
\ee
This indeed vanishes for the conformal plasmas with $X=0$.

\subsection{Bjorken flow in the CR background} \label{sec:CRBj}

Let us now construct the late time behavior of the Bjorken flow for the CR solution discussed above, following closely the analysis of~\cite{Janik:2005zt}. 
We start from the zero temperature solution so that $C_2=0$ in~\eqref{fsol} and the 
blackening factor $f$ is identically equal to one. We use the metric~\eqref{conf} in the conformal coordinate system, denoting 
 $z = r/\ell'$. For simplicity we also set $A_0=0$ and $\ell'=1$; 
the metric becomes
\be\lab{dz} 
ds^2 = z^{-\frac{2}{1-4X^2}} \le(dz^2 - dt^2 + \delta_{ij} dx^{i} dx^{j} \ri)  \, ,
\ee
and the dilaton solution reads
\be\label{dilz}
 \l = z^{-\frac{3X }{1-4 X^2}} 
\ee
when the potential is normalized as 
\be\label{normpot}
 V(\l) =\frac{12(1-X^2)}{(1-4X^2)^2} \l^{-\frac{8 X}{3}} \, .
\ee
In order to study the Bjorken flow, we switch to the proper time $\t$ and pseudo rapidity $y$ as in~\eqref{coors}.
Following~\cite{Janik:2005zt}, we define the scaling variable
\be
 v = \frac{z}{\t^{s/4}}\, ,
\ee
where $0<s<4$. We then study a ``variation'' of the metric~\eqref{dz} at late times, $\t \to \infty$, keeping $v$ fixed. 
We first write an Ansatz for the metric in a form where the gauge has not yet been fixed: 
\be \label{dabc}
 ds^2 = z^{-\frac{2}{1-4X^2}} \le(e^{d(v)}dz^2 - e^{a(v)}d\t^2 + e^{b(v)}\t^2 dy^2 + e^{c(v)} dx_\perp^2 \ri) \, .
\ee
As the CR solution contains a nontrivial dilaton profile, we must allow for it to vary as well. Therefore we write
\be \label{laansatz}
 \l = z^{-\frac{3X }{1-4 X^2}} e^{\l_1(v)} \, .
\ee
This Ansatz can then be substituted in the equations of motion
\bea \label{Einsteqs}
 R^\mu_{\ \nu}- \frac{1}{2} R\, \delta^\mu_{\ \nu} &=& \frac{4}{3\l^2}\left[\pa^\mu \l\, \pa_\nu\l -\frac{1}{2} \left(\pa \l\right)^2 \delta^\mu_{\ \nu}\right] + \frac{1}{2} \delta^\mu_{\ \nu} V(\l) \\
 \label{dileq}
 \Box \l &=& - \frac{8}{3} \l\, V'(\l) \, .
\eea
Interestingly, a simple special solution to the time dependent problem can be found quite easily. 
Namely, we can follow the arguments in~\cite{Bak:2006dn}: The diagonal components of the Einstein equations for the static geometry
\be \label{dstatic}
 ds^2 = z^{-\frac{2}{1-4X^2}} \le(e^{d(z)}dz^2 - e^{a(z)}dt^2 + e^{b(z)} dx_1^2 + e^{c(z)} dx_\perp^2 \ri) 
\ee
(and replacing $\l_1(v)$ by $\l_1(z)$ in~\eqref{laansatz}) 
have the same form as the diagonal components of the Einstein equations for the evolving metric~\eqref{dabc} at leading order in $1/\t$. In particular,
the evolving diagonal Einstein equations are solved at leading order in $1/\t$ by the analogue of the static BH solution in~\eqref{CRconf1}:
\be \label{specialsol}
 a(v) = - d(v) = \log\left[1-\left(\frac{v}{v_h}\right)^{\frac{4(1-X^2)}{1-4X^2}}\right]\, , \qquad b(v)=c(v)=\l_1(v)=0 \, ,
\ee
for any $s$ with $0<s<4$. The nondiagonal Einstein equation (the $z\t$ component) yields the additional equation
\be \label{seq}
\left(3 s-4+16 X^2\right) a'(v)-(s-4) \left(1-4 X^2\right) b'(v)-2 s \left(1-4 X^2\right) c'(v)-8 s X \l_1'(v) = 0 \, .
\ee
This is the only equation which depends on $s$ explicitly, and it is also satisfied if
\be
 s = \frac{4}{3}(1-4X^2) \, .
\ee
The dilaton equation of motion~\eqref{dileq} follows from the Einstein equations and is therefore also automatically satisfied. 
We will next show that~\eqref{specialsol} is actually the only nontrivial solution which has regular behavior in the IR.

\subsubsection{General analytic solution}

Let us then fix the gauge $d=0$ and study for the ``variation'' $(a,b,c,\l_1)$. 
At leading order in $1/\t$ the Einstein equations yield a nonlinear system of equations given as~\eqref{nleq1}--\eqref{nleq5} 
(and including~\eqref{seq}) in Appendix~\ref{App-Gensol}. 
Remarkably, the general solution to this system can be found analytically. 

First it is useful to do a change the basis of functions which generalizes~\eqref{CRchange} to finite $X$:
\bea
 \label{basis1}
 a(v) &=& A(v)-2\left(1-4 X^2\right) m(v)+2 X n(v) \\
 b(v) &=&  A(v)+2 \left(s-1+4 X^2\right)m(v) +2 X n(v) \\
 c(v) &=&  A(v)- \left(s-2+8 X^2\right)m(v)-2 X n(v) \\
 \label{basis4}
 \l_1(v) &=&  \frac{3}{2} X A(v)+X\left(1-4 X^2\right) m(v)+\left(1-X^2\right) n(v) \, .
\eea
Notice that the coefficients were chosen such that~\eqref{seq} is automatically satisfied. By taking suitable combinations of the other Einstein equations we obtain %The other equations can then be combined to give
\be
m''(w) = \xi m'(w)-2 A'(w) m'(w)\,, \qquad \frac{m''(w)}{m'(w)} = \frac{n''(w)}{n'(w)} \, ,
\ee
where  $w = \log v$ and $\xi$ is defined in (\ref{xi}). 
From here one readily obtains
\be \label{Ansol}
 A(w) = \frac{\xi}{2}w - \frac{1}{2} \log m'(w) + \mathrm{const.}\, , \qquad n(w) = \k\, m(w) + \mathrm{const.}\,,
\ee
where 
the integration constant $\k$ can take any real value.
Therefore $A$ and $n$ can be eliminated from the system of equations. The remaining single equation can be written in a polynomial form by using the derivative $p(w)=m'(w)$:
\begin{align} \label{peq}
&\frac{8 \left(1-X^2\right) X^4}{\left(1-4 X^2\right)^2}+4 \frac{K  X^2}{1-4X^2} p(w)-\frac{\S^2-K ^2}{2 \left(1-X^2\right)}p(w)^2
\\\nn
+&K p'(w)+ \frac{2 X^2-4 X^4}{1-4 X^2 }\frac{p'(w)}{p(w)}-\frac{1+X^2 }{2 }\frac{p'(w)^2}{p(w)^2}
=-\frac{p''(w)}{2 p(w)}\,,
\end{align}
where
\bea \label{Sdef}
 \S &=& \frac{4}{3}\sqrt{\left(1-4 X^2\right)^2+\frac{1}{8}\left(1-X^2\right) \left(3 s-4+16 X^2+4 \kappa  X\right)^2 + \left(1-X^2\right)\kappa ^2 }\,,\mbox{\phantom{aaaaa}} \nonumber \\
 K &=& \frac{4 }{3 } X\left(X-4 X^3+\kappa -\kappa  X^2\right)\,.
\eea

The general solution to~\eqref{peq} is discussed in Appendix~\ref{App-Gensol}. The solution which is regular in the UV, i.e. it has an analytic expansion in the variable
\be
 v^{\xi } = e^{\xi\,w}
\ee
can be written as
\begin{align} \label{analyticsol}
 &w = \log v = -\frac{\S+K }{\xi X^2 }m  +\frac{1}{\xi} \log \left(e^{2 \S m}-1\right)\nn\\
 -&\frac{1-4X^2}{4 X^2} \log\, _2F_1\left(1,\frac{\S(1-2 X^2)+K}{2 \S(1-X^2)};\frac{1-2 X^2}{1-X^2};1-e^{2 \S m}\right)\, .
\end{align}
That is, the inverse function $w(m)$ could be found in closed form. This solution has a ``horizon'' at a finite value of $w$ where $m$ tends to infinity, which screens the IR singularity at $w=+\infty$. 
Therefore $w$ runs from $-\infty$ in the UV to a finite value in the IR, whereas $m$ runs from zero to $+\infty$.

The metric~\eqref{dabc} takes a rather simple form when evaluated on the analytic solution. As the result~\eqref{analyticsol} suggests, it is convenient to use $m$ as the bulk coordinate instead of $v$. 
We also fix the constant terms appearing in~\eqref{Ansol} such that $A$ and $n$ vanish in the UV. %, and set $w_0=0$ in~\eqref{analyticsol}. 
Inserting the result in~\eqref{dabc}, we obtain
\bea \label{metricgeneral}
 ds^2 &\simeq& \t^{-\frac{s}{2(1-4X^2)}}\bigg\{\t^{s/2}\le(\frac{2\S}{\xi}\ri)^2\le(1-e^{-2 \S m}\ri)^{-\frac{2}{1-X^2}}e^{\frac{-2\S+2K}{1-X^2} m}\, dm^2\nn\\
 &&+\le(1-e^{-2 \S m}\ri)^{-\frac{1}{2(1-X^2)}}\bigg[- e^{\frac{-\S+4 K}{2 \left(1-X^2\right)}m}e^{-\frac{8 m}{\xi}}d\t^2 \\\nn
&&+ \t^2 e^{\frac{-\S+4K}{2 \left(1-X^2\right)}m} e^{ \le(2s-\frac{8}{\xi}\ri) m} dy^2+e^{\frac{-\S-2K}{2 \left(1-X^2\right)}m} e^{\le(\frac{8}{\xi}- s\ri) m} dx_\perp^2\bigg] \bigg\}\,,
\eea
where we dropped the nondiagonal terms as well as a term in the ${\t\t}$-component, which are irrelevant for the dynamics at leading order in $1/\t$. Interestingly, the hypergeometric function appearing in the solution~\eqref{analyticsol} cancels in the metric 
so that the leading terms can be expressed in terms of elementary functions. Similarly, the dilaton can be written as
\be
 \l = \t^{-\frac{3 s X}{4(1-4X^2)}} \le(1-e^{-2 \S m}\ri)^{-\frac{3X}{4(1-X^2)}}e^{\frac{3 \left(-\S X^2+K\right)}{4 X \left(1-X^2\right)}m}\,.
\ee
Notice that the constant factor $2\S/\xi$ in the first line of~\eqref{metricgeneral} could be eliminated by varying the value of the constant $\ell'$ (which was set to one above).

\subsubsection{IR regularity}

Let us then analyze the behavior of the metric in the IR, $m \to +\infty$. We first define
\be
 \rho = \exp\le[ \frac{-\S+K}{1-X^2} m\ri] \, .
\ee
It is straightforward to show that $\rho \to 0$ in the IR for all allowed values of $X$, $s$, and $\k$. Changing variables from $m$ to $\rho$, the various components of the metric have the behavior
\begin{align} \label{ghorizon}
 g_{\rho\rho} &\sim \rho^0\, ,&\qquad g_{\t\t} &\sim \rho^\frac{\S-4 K+4 \left(1-4 X^2\right)}{2 \left(\S-K \right)}\,,&\\\nn
 g_{yy} &\sim \rho^\frac{\S-4K-4 s \left(1-X^2\right)+4 \left(1-4 X^2\right) }{2 \left(\S-K  \right)}\,,& 
 g_{xx} &\sim \rho^\frac{\S+2K+2 s \left(1-X^2\right)-4 \left(1-4 X^2\right)}{2 \left(\S-K \right)}\,,&
\end{align}
as $\rho \to 0$, where $g_{xx}$ stands for the transverse components of the metric. 
For the static black hole metric, after a similar change of variables the component $g_{\t\t}$ is $\propto \rho^{2}$ while the other components take constant values as $\rho \to 0$~\cite{Janik:2005zt}. 
Recalling the definitions~\eqref{Sdef}, it is not difficult to see that the same holds for the evolving metric only if 
\be \label{IRregconds}
 s = \frac{4}{3}\le(1-4X^2\ri)\,,\qquad \k = 0 \,,
\ee
so that $\S = 4(1-4X^2)/3$ and $K=4X^2(1-4X^2)/3$. We note that,  the latter condition means one of the fluctuation mode decouples by equation (\ref{Ansol}).  
We have also verified numerically, by computing the Ricci scalar and the squared Riemann tensor as $m\to\infty$, that a curvature singularity is only avoided when the conditions~\eqref{IRregconds} hold. 

Substituting the conditions~\eqref{IRregconds} in the general expression~\eqref{metricgeneral}, 
we obtain an explicit formula for the regular metric:
\bea \label{metricregular}
 ds^2 &\simeq& \t^{-\frac{8}{3}X^2}\le[\frac{8(1-4 X^2)}{3\xi}\ri]^2\le(1-e^{-\frac{8}{3}(1-4X^2) m}\ri)^{-\frac{2}{1-X^2}}e^{-\frac{8}{3}(1-4X^2) m}\, dm^2\\\nn
 &&+\t^{-\frac{2}{3}}\le(1-e^{-\frac{8}{3}(1-4X^2) m}\ri)^{-\frac{1}{2(1-X^2)}} 
\Big[\!- e^{-\frac{8}{3}(1-4 X^2) m}d\t^2+ \t^2  dy^2+ dx_\perp^2 \Big]\,,
\eea
and the dilaton solution becomes
\be
 \l =  \t^{-X} \le(1-e^{-\frac{8}{3}(1-4X^2)m}\ri)^{-\frac{3X}{4(1-X^2)}}\,.
\ee
After the change of variables\footnote{Notice that the definition of the scaling variable depends on the gauge. 
Explicitly,~\eqref{analyticsol} gives the definition in the $d=0$ gauge whereas~\eqref{hatvdef} is the definition in, e.g., $c=0$ gauge.}
\be \label{hatvdef}
 \hat v =  \le(1-e^{-\frac{8}{3}(1-4X^2)m}\ri)^\frac 1\xi
\ee
we find that the solution takes a form which is consistent with~\eqref{specialsol} (up to subleading terms in $1/\t$):
\bea \label{simpleregular}
 ds^2 &=& \hat v^{-\frac{2}{1-4X^2}}\le\{  \frac{\t^{-\frac{8}{3}X^2} }{1-\hat v^\xi}d\hat v^2 + \t^{-\frac{2}{3}} \le[-\le(1-\hat v^\xi\ri)d\t^2+ \t^2  dy^2+ dx_\perp^2 \ri]\ri\} \\
 \l &=&  \t^{-X} \hat v^{-\frac{3X}{1-4X^2}} \, .
\eea
We may further identify
\be \label{zczerog}
 \hat z = \hat v\, \t^{s/4} = \hat v\, \t^\frac{1-4X^2}{3}
\ee
so that the solution becomes that of black hole with a moving horizon: 
\bea \label{simpleregular2}
 ds^2 &\simeq& \hat z^{-\frac{2}{1-4X^2}}\le\{  \frac{d\hat z^2}{1 - \t^{-\frac{4}{3}(1-X^2)} \hat z^\xi} -\le(1-\t^{-\frac{4}{3}(1-X^2)} \hat z^\xi\ri)d\t^2+ \t^2  dy^2+ dx_\perp^2\ri\} \nn\\
 \l &=&  \hat z^{-\frac{3X}{1-4X^2}} \, ,
\eea
where we again dropped the nondiagonal terms in the metric as well as an extra term in the $\t\t$ component, 
which are irrelevant since they do not enter the dynamics (i.e., the Einstein equations) at leading order in $1/\t$.

We can then also confirm that the Ricci scalar and the squared Riemann tensor are indeed regular for this metric in the IR similarly as in the conformal case of~\cite{Janik:2005zt}:
\bea \label{Rvalue}
 R &\simeq& - 20 \frac{1-X^2}{(1-4X^2)^2} \t^{8 X^2/3} \\
 \label{RRvalue}
 \mathfrak{R}^2 &=& R^{\m\n\a\b}R_{\m\n\a\b} \simeq 112 \frac{(1-X^2)^2}{(1-4X^2)^4} \t^{16 X^2/3}
\eea
up to corrections suppressed by $1/\t$ or by $\exp(- m)$ (or equivalently by $(1-\hat v)$). Since the values of these scalars increase with $\t$, it is essential to first consider the leading solution in
$1/\t$ and impose its regularity on the horizon. 
For the general solution  of~\eqref{analyticsol} the expressions for $R$ and $\mathfrak{R}^2$ become rather complicated, 
but as we mentioned above, we have verified numerically that all other choices except for those given in~\eqref{IRregconds} lead to a curvature singularity at the horizon.

\subsection{Continuation of the result to $-1<X\leq -1/2$ and thermodynamics}

The final regular metric in~\eqref{simpleregular2} appears singular at $X=-1/2$ where $\xi$ also diverges. It is, however, quite easy to absorb the singularity
by a suitable redefinition of coordinates and variables: recall from above that such a singularity was absent for the static BH in the domain wall coordinates but appeared after the change 
to conformal coordinates in~\eqref{CRconf1}. First we reinstate the dependence of the metric and the dilaton potential on $\ell'$:
\bea
 ds^2 &=&  \hat z^{-\frac{2}{1-4X^2}}\le\{  \frac{(\ell')^2d\hat z^2}{1 - \t^{-\frac{4}{3}(1-X^2)} \hat z^\xi} -\le(1-\t^{-\frac{4}{3}(1-X^2)} \hat z^\xi\ri)d\t^2+ \t^2  dy^2+ dx_\perp^2\ri\}\nn \\
 V &=& \frac{12(1-X^2)}{(1-4X^2)^2(\ell')^2} \l^{-\frac{8 X}{3}} \ .
\eea
By setting here
%\be
$\ell' = 1/(1-4X^2)$,
%\ee
which corresponds to $\ell = e^{A_0}$ in~\eqref{CRconf1}, the divergence in the dilaton potential is cancelled. Next we switch to an analog of the domain wall coordinates, 
\be
 \hat z = \le(-4X^2 \hat u\ri)^{-\frac{1-4X^2}{4 X^2}} \,.
\ee
The resulting metric and the dilaton potential read
\bea
ds^2 &=&  \le(1-\t^{-\frac{4}{3}(1-X^2)}(-4X^2\hat {u})^{-\frac{1-X^2}{X^2}}\ri)^{-1}d\hat u^2\\\nn
&&+\le(-4X^2\hat u\ri)^{\frac{1}{2X^2}}
    \le\{ -    
\le(1-\t^{-\frac{4}{3}(1-X^2)}(-4X^2\hat u)^{-\frac{1-X^2}{X^2}}\ri)d\tau^2 + \t^2  dy^2+ dx_\perp^2\ri\}\\
 \l &=&  \le(-4X^2\hat u\ri)^\frac{3}{4X} \\\nn
 V &=& 12(1-X^2) \l^{-\frac{8 X}{3}}
\eea
so that the singularity at $X=-1/2$ has indeed been removed, and the metric is a boost invariant version of the static black hole~\eqref{BHmet} with $A_0=0=C_1$, $\ell=1$, and with a time-dependent $C_2$.

Finally let us comment on the thermodynamics of this solution. It is tempting to simply apply the formula~\eqref{temp} and compute the entropy from the size of the shrinking black hole, 
even if it is not obvious that this is a valid procedure for an evolving system. The obtained temperature and entropy density (for the volume element $\tau dy dx_2 dx_3$) read
\be \label{TSfromBH}
 T = \frac{1-X^2}{\pi} \tau^{-\frac{1}{3}(1-4X^2)}\, , \qquad \  S = \frac{1}{4G_5}e^{3 A_h} = \frac{1}{4 G_5\, \tau}\, ,
\ee
where $A_h$ is the value of the scale factor $A$ at the horizon. These results imply the energy  and free energy densities
\be
 \epsilon = \frac{3}{16 \pi G_5} \tau^{-\frac{4}{3}(1-X^2)} \,,\qquad F = -\frac{1-4X^2}{16 \pi G_5} \tau^{-\frac{4}{3}(1-X^2)} \, .
\ee
Notice that these formulas agree (if $\tau$ is eliminated) with the thermodynamics of the static CR plasma given in section~\ref{sec:CRthermo}. 
For the decay of the energy density in~\eqref{solepar}, $\epsilon \sim \tau^{-\alpha \xi}$, we find
\be\lab{num} 
\alpha \xi = \frac43 (1- X^2)\, .
\ee
This result will be confirmed by computing the energy-momentum tensor from the boundary data below. 
Comparing with the definition (\ref{solepar}) for the temperature behavior we find the relation between the parameters $\alpha$ and $s$: 
\be\lab{lateT} 
 \alpha = \frac{s}{4} = \frac13 (1-4X^2)\, ,
\ee
which reduces to  $T\sim \tau^{-1/3}$ in the conformal case $X=0$. We also see that the condition (\ref{req}) is satisfied. 
We see again that the free energy of the evolving solution is positive for $-1 < X -1/2$, indicating that the solution is unstable.
As another remark, the fact that the entropy density in~\eqref{TSfromBH} is inversely
proportional to $\tau$ is expected for perfect fluid, since there is no entropy production and the volume of the plasma is proportional to $\tau$ for Bjorken flow.

\subsection{Holographic stress-energy tensor} \label{sec:EMtensor}

We can also check the results of the thermodynamics of the evolving solution by an explicit holographic computation of the renormalized boundary stress-energy tensor. The most efficient way to extract the stress-energy tensor of the dual theory is to lift the CR solution to a higher dimension where it becomes asymptotically AdS. 
We consider the diagonal reduction as in section (2.1) of \cite{Gouteraux:2011qh} in the case where the internal manifold is flat. Let us review the procedure. Starting from the higher dimensional action

\be\label{highdim}
S = \frac{1}{16 \pi \tilde G_N} \int \, d^{d+1}x \, d^{2\sigma-d} y \, \sqrt{-\tilde g} \left( \tilde R - 2 \Lambda \right) 
\ee  
and using the following Ansatz for the metric on ${\cal M}^{d+1} \times {\mathbb R}^{2\sigma-d}$
\be\label{liftmetric}
\widetilde{ds}^2 = e^{- \delta_1\phi(x)} dx^2 + e^{\delta_2 \phi(x)} dy^2
\ee
we find 
\begin{eqnarray}
e^{-\delta_1 \phi}\tilde R &=& R + \left( \delta_1 d + \delta_2 (d-2 \sigma) \right) \, \nabla \cdot \partial \phi + \nonumber \\
& + & \left[ - {d (d-1) \over 4} (\delta_1+\delta_2)^2 +  \sigma(d-1) ( \delta_1+\delta_2 )\delta_2 - 
  {\sigma (2 \sigma-1) \over 2} \, \delta_2^2
\right]\,  (\partial \phi)^2  \nonumber \,.
\end{eqnarray}
Requiring that the final action is in the Einstein frame and the dilaton is canonically normalized (so that is has a factor $4/(d-1)$ in the kinetic term) we have   
\be
\delta_1 = { 4 \sqrt{2\sigma - d} \over (d-1) \sqrt{(2 \sigma-1)} } \,, \quad \delta_2 =  
 { 4  \over \sqrt{ (2 \sigma-1)(2 \sigma-d) } } \,.
\ee
The dilation potential comes from the cosmological constant in (\ref{highdim});  requiring it to be $V_0 e^{-8 X\phi/3}$ gives 
$$2 \sigma - d = { 4 (d-1)^3 X^2 \over 9 -4 (d-1)^2 X^2 } \,. $$ 
The number of extra dimensions goes from 0 to $\infty$ for $X \in [-1/2,0]$ (in $d=4$), so the number of counterterms required to regularise the action depends on the value of $X$. However,  since the uplifted metric (\ref{liftmetric}) is asymptotically AdS we can read off the energy momentum tensor simply from the appropriate coefficient of the metric in the Fefferman-Graham expansion: 
\be 
\langle T^{\mu\nu} \rangle_{2\sigma} = {2 \sigma l^{2\sigma-1} \over 16 \pi \tilde G_N } \tilde g^{\mu\nu}_{(2 \sigma)} \,,
\ee
where $l$ is the AdS radius: $\Lambda = - \sigma(2\sigma-1)/l^2$.
To obtain the $d$-dimensional tensor we need to take into account the (infinite) volume of the compactification manifold, that we reabsorb in a redefinition of the Newton's constant, and the rescaling of the induced metric on the boundary. 
For $d=4$ we have
$$\sigma = 2 { 1-X^2 \over 1-4 X^2} \,, \quad \delta_1 = { 8 X \over 3}\,, \quad \delta_2 = - {2\over 3X} (1-4X^2) \,.$$
The uplifted metric is 
\begin{equation}
\tilde g \equiv  {d\psi^2 \over \psi^2} + {1\over \psi^2} (\tilde \gamma^{(0)} + \ldots ) = \lambda^{-\delta_1} z^{-{2\over 1-4 X^2}} (dz^2 + \gamma^{(0)} ) + \lambda^{\delta_2} dy^2 \,, 
\end{equation}
where the first equation defines the Fefferman-Graham coordinate $\psi$. Using the dilaton (\ref{dilz}) one can show that we can simply identify $\psi = z,  \gamma^{(0)}_{\mu\nu} = \tilde \gamma^{(0)}_{\mu\nu}, \sqrt{\gamma^{(0)}} = \sqrt{\tilde \gamma^{(0)}}$. 
This implies 
\be
\langle T^{\mu\nu} \rangle_d = {1 \over \sqrt{\gamma^{(0)}}} {\delta S \over \delta \gamma^{(0)}_{\mu\nu} } = 
 \langle T^{\mu\nu} \rangle_{2\sigma}  \,.
\ee
  
 We can also carry out the holographic renormalization, using the results found in \cite{Papadimitriou:2011qb} that apply to a generic Einstein-dilaton theory. In this paper we are considering only the leading order in the derivative expansion, and the renormalization is straightforward: the only counterterm needed is 
 \be
 S_{ct} = - \frac{1}{8 \pi G_5}\int_{\partial M} \sqrt{\gamma} \, U(\phi) \,,
 \ee
 where $U$ is the superpotential, related to the dilaton potential by 
 \be\label{superpot}
  \frac{3}{4} U'(\phi)^2 -   \frac{4}{3} U(\phi)^2 + V(\phi)=0 \,.
  \ee
With our choice of potential (\ref{normpot}) it is given by $$U = \frac{3}{1-4X^2} e^{-{4 \over 3} X \phi} \,.$$
 
Even though it is not a priori obvious that the renormalization procedure should commute with the generalised dimensional reduction, and it has not been proven that it does in general \cite{papadim-private}, in this case we find agreement between the two procedures; the renormalized energy-momentum tensor is 
\be
T_{\mu\nu} = t_0 (1-4X^2) \tau^{-{4\over 3}(1-X^2)}   \,  \mathrm{diag}(-{3 \over 1-4X^2}, \tau^2,1,1 ) \,,
\ee
consistent with the field theory expectation (\ref{emtCR},\ref{solepar},\ref{num}). Here $t_0$ is an arbitrary parameter proportional to the integration constant appearing in the UV expansion of the solution (\ref{analyticsol}), which can be found in Appendix~\ref{App-Gensol}. 
Notice that in the limit $X \to -1/2$ the stress-energy tensor remains finite and it reduces to a pressureless gas. 

\section{Modified CR} \label{CRpow}
\subsection{IR modified black brane solution} \label{sec:IRmod}

We also consider the black-brane solution of a single exponential potential modified with power-law scaling: 
\begin{equation}\label{exppotmod}
    V  = V_1\,e^{-\frac83 X_0\, \phi}\phi^{P}\, ,
\end{equation}
where $-1<X_0<0$, $P>0$ and $V_1>0$ are real constants \footnote{We denote the exponent by $X_0$ because in this modified potential it can no longer be identified with the constant value of the variable $X(\phi)$ defined in Appendix \ref{App-modsol}.}. These type of potentials, in particular for $X_0=-1/2$, $P=1/2$, are singled out in the improved holographic QCD program \cite{ihqcd1,ihqcd2,ihqcd3,ihqcd4} as the large dilaton limit of a choice of theories that yield the best fit to the glueball spectra and thermodynamics \cite{ihqcd5}. Notice that this form of the potential can be valid only in the IR, since it is not well-defined when $\phi < 0$.   

The solution to (\ref{exppotmod}) cannot be obtained analytically. However, we are interested in the IR limit\footnote{This limit corresponds to the late-time behavior of the non-static black brane solution, as discussed in the next section.} where large $\phi$ approximation can be used to construct an analytic solution given in powers of $1/\phi$.
We are interested in the black brane solutions with a horizon $u_h$ on which the value of the dilation is $\phi_h$. The solution presented below will be valid in the limit $1\ll \phi \ll \phi_h$. The derivation is presented in Appendix \ref{App-modsol}. The solution is most easily expressed in a coordinate system where $\phi$ is the radial variable, 
\be\lab{df} 
ds^2 = e^{2A(\phi)} \le(-f(\phi) dt^2 + \delta_{ij} dx^{i} dx^{j} \ri) + e^{2B(\phi)}\frac{d\phi^2}{f(\phi)}, \qquad \phi=\phi(u)\, ,
\ee
where,
\bea
\lab{Af} 
e^{2A(\phi)} &=& e^{2A_0} e^{2\phi / (3X_0)} \phi^{P/(4X_0^2)} \le(1 + \cO(1/\phi)\ri)\, ,\\
\lab{Bf}
e^{2B(\phi)} &=& e^{2B_0} e^{\frac83X_0\phi} \phi^{-P} \le(1 + \cO(1/\phi)\ri)\, ,\\
\lab{ff}
f(\phi)  &=& 1 - e^{a(\phi_h-\phi)} (\phi_h/\phi)^b\le(1 + \cO(1/\phi) \ri)\, .
\eea
Here $A_0$ is an integration constants, $\phi_h$ is the location of the horizon and 
\be\lab{cons}
e^{2B_0} = \frac{4(1-X_0^2)}{3 X_0^2 V_1}, \qquad a = \frac{4(1-X_0^2)}{3X_0},\qquad b = \frac{P(1 + X_0^{-2})}{2}\, ,
\ee
are constants. The $\cO(1/\phi)$ corrections can be determined analytically order by order. 
This solution reduces to (\ref{Asol} and (\ref{fsol})  when $P=0$.  
Conversion to domain-wall coordinates is straightforward. One finds 
\be\lab{phiumod}
\lambda = e^{\phi} \approx ( 4X_0 c(u-u_0)/3)^{3/(4X_0)} ( 3/(4X_0) \log(4X_0 c (u-u_0)/3 ))^{3P/(8X_0)}\, , 
\ee
where $c$ is a constant, see Appendix \ref{App-modsol}.

Thermodynamics of the IR modified potential %in section \ref{sec:IRmod} 
can also be obtained analytically. The entropy and the temperature as functions of the horizon value of the scalar field $\phi_h$ are given by 
\bea\lab{sIRmod}
  S &=& \frac{e^{3A_0}}{4G_5}e^{\frac{\phi_h}{X_0}}\phi_h^{\frac{3P}{8X_0^2}}\, ,\\
\lab{TIRmod}
 T &=& - e^{A_0-B_0}\le(a + \frac{b}{\phi_h}\ri) e^{\le(\frac{1}{X_0}-4\frac{X_0}{3}\ri)\phi_h }\phi_h^{\frac{P}{2}\le(1+\frac{1}{4X_0^2}\ri)}\, .
 \eea
One can obtain $S$  
as a function of $T$ by parametrically solving those two equations in $\phi_h$.  Then the calculation of the free energy and the energy follows as in the case $P=0$ above.

The bulk viscosity of theory with the IR modified potential
can also be calculated. For this calculation it is more convenient to use the analytic expression for the bulk viscosity for an arbitrary potential $V(\phi)$ that was first obtained in \cite{ElingOz}: 
\be\lab{bv2} 
\frac{\zeta}{S}  = \sqrt{\frac83}\frac{S}{4\pi} \frac{d\phi_h}{dS}\, .
\ee
Substitution of (\ref{sIRmod}) into this expression yields, 
\be\lab{bv3} 
\frac{\zeta}{S}  = \frac{1}{4\pi}\sqrt{\frac83} \frac{X_0}{1+\frac{3P}{X_0\phi_h}}\, , 
\ee
which indeed reduces to (\ref{bv1}) for $P=0$. Finally one can express this in terms of temperature by inverting the formula (\ref{TIRmod}). The difference between the
IR non-modified and the modified cases is then that in the former case it does not depend on $T$, whereas it does in the latter case.

\subsection{Evolving metric in the presence of an IR modification} 

Let us then discuss how the evolving metric changes if the dilaton potential is modified by a power-law function in the IR as above in~\eqref{exppotmod}.
We will restrict to finding the generalization of the special solution~\eqref{specialsol}. As in section~\ref{sec:IRmod}, we choose to use the dilaton as the radial coordinate, 
and write an Ansatz which has a form similar to the static metric:
\be \label{evolvingansatz}
 ds^2 = e^{2A(\phi)} \le(-f(\bar w,\log \tau) d\tau^2 + \tau^2dy^2 + dx_\perp^2 \ri) + e^{2B(\phi)}\frac{d\phi^2}{f(\bar w,\log \tau)}
\ee
where the functions $A(\phi)$ and $B(\phi)$ are exactly the same functions as in the static case, and  $\bar w$ is a scaling variable analogous to $w$ of the previous section, which will be specified below. 
As it turns out, the computation does not proceed exactly as in~\cite{Bak:2006dn} (or as reviewed in Sec.~\ref{sec:CRBj}):
the scaling variable cannot be chosen such that the evolving blackening factor $f(\bar w)$ would satisfy the same equations as the static blackening factor $f(\phi)$ at leading order in $1/\tau$. 
Instead, we can solve the Einstein equations for $f$ as series in $1/\log \tau$ (so that corrections in $1/\tau$ are highly suppressed) keeping a well chosen scaling variable $\bar w$ fixed. This is why we also included $\tau$ dependence in the blackening
factor in the Ansatz~\eqref{evolvingansatz}. A good choice for $\bar w$ turns out to be
\be\label{scalingvar}
 \bar w = \phi + s_1 \log \tau + s_2 \log \log \tau
\ee
which will take a fixed value at the horizon for a convenient choice of the coefficients $s_i$ as $\tau$ increases.

At leading order in $1/\tau$, the $\phi\phi$ and $\phi\tau$ components of the Einstein equations imply for the blackening factor
\bea \label{Einst1}
 9A'(\phi)\frac{\pa f(\bar w,\log \tau)}{\pa \bar w} &=& -4(1-f(\bar w,\log \tau))\le(1-9A'(\phi)^2\ri) \\
 \label{Einst2}
 3A'(\phi)\frac{\pa f(\bar w,\log \tau)}{\pa \log \tau} &=&   \left[1-3 \le(s_1 +\frac{s_2}{\log\tau}\ri)A'(\phi)\right]\frac{\pa f(\bar w,\log \tau)}{\pa \bar w}
\eea
where we also used the fact that $A(\phi)$ and $B(\phi)$ satisfy the static Einstein equations. Inserting the asymptotic expansion for $A(\phi)$ from~\eqref{Af} and developing as a series at large $\log\tau$, the first equation~\eqref{Einst1} is solved by
\be
 f(\bar w,\log \tau) = 1 - f_0\, e^{-4\bar w(1-X_0^2)/(3 X_0)} + \mathcal{O}\le(\frac{1}{\log\tau}\ri)
\ee
which also solves the second equation~\eqref{Einst2} up to next-to-leading order if we choose
\be
 s_1 = X_0 \,,\qquad s_2 = \frac{3 P}{8 X_0}\, .
\ee
Therefore the leading order solution for the blackening factor in terms of $\phi$ and $\tau$ reads
\be\label{fasympmodpot}
 f \simeq 1 - f_0\, \tau^{-4(1-X_0^2)/3}\, \le(\log\tau\ri)^{{-P(1-X_0^2)/(2 X_0^2)}}\, e^{-4\phi(1-X_0^2)/(3 X_0)}\,.
\ee
Notice that the form of the blackening factor is similar to that of the static solution in~\eqref{ff}, but the exponent of the logarithmic term involving $P$ is slightly different. 
Naturally, the solution also agrees with that obtained without the IR correction in the limit $P \to 0$, see~\eqref{simpleregular2}. It is possible to find the solution at higher orders in $1/\log\tau$ if one writes
a more generic Ansatz for the metric than that of~\eqref{evolvingansatz}, in analogy to~\eqref{dabc} above.

Expected scaling laws for the thermodynamics can be extracted from the evolving metric in the same way as in the absence of logarithmic corrections in the previous section. That is, if we apply the formulae of the static solution, we find that
\bea
 T &=& \frac{1}{4 \pi} |\partial_\phi f| e^{A_h-B_h} \simeq \frac{(1-X_0^2)}{\pi} (-X_0)^{\frac{P}{8X_0^2}(1+4X_0^2)} \tau^{-\frac{1}{3}(1-4X_0^2)} (\log\tau)^P\,, \\
 S &=& \frac{1}{4 G_5} e^{3 A_h} \simeq \frac{(-X_0)^\frac{3 P}{8X_0^2}}{4 G_5 \tau}\,,
\eea
where we set $A_0=0$ and $V_1 =12(1-X_0^2)$ to be able to compare to~\eqref{TSfromBH} directly. 
Notice that the time dependence arises solely from the scale factors $A_h$ and $B_h$ evaluated at the horizon.  
Interestingly, the logarithmic corrections to the entropy density cancel. 
Therefore, finite value of $P$ does not lead to production of entropy, in agreement with the perfect fluid picture.  
If $X_0=-1/2$, temperature decreases with $\tau$ only for $P<0$.

The corresponding energy density and pressure read
\bea
 \epsilon &\simeq& \frac{3(-X_0)^{\frac{P}{2X_0^2}(1+X_0^2)}}{16 \pi G_5} \tau^{-\frac{4}{3}(1-X_0^2)}(\log\tau)^P \,,\\
 F =  -p&\simeq& -\frac{1-4X_0^2}{16 \pi G_5}(-X_0)^{\frac{P}{2X_0^2}(1+X_0^2)} \tau^{-\frac{4}{3}(1-X_0^2)}(\log\tau)^P \, .
\eea
For  $X_0 = -1/2$ the above expression for the free energy vanishes. In this case the leading nonzero expression for the free energy
is suppressed by $1/\log \tau$:
\be
 F \simeq \frac{3 P}{16  \pi G_5 }2^{-P} (\log \tau)^{P-1} \, .
\ee
The critical value where the free energy changes sign is therefore $P=0$, which is the same value where the static configurations change from confining to deconfining~\cite{ihqcd1,ihqcd2}, 
in analogy to the value $X_0 = -1/2$ for the leading power behavior.

\subsection{Holographic stress-energy tensor} 

The power-law corrected potential (\ref{exppotmod}) does not come from a generalized dimensional reduction. We also  cannot carry out the holographic renormalisation reliably, as we do not know the asymptotic form of the solution in the UV. Nevertheless, if we attempt to extract the finite, T-dependent part of the stress-energy tensor as in Sec.~\ref{sec:EMtensor} using the expansion (\ref{fasympmodpot}), we find 
\begin{eqnarray}
\epsilon &\sim&  \frac{3 f_0}{2(1-4X_0^2)} \phi ^{\frac{P}{2}(1+\frac{1}{X_0^2})} \tau^{-\frac{1}{3}(1-4X_0^2)} (\log \tau)^{\frac{P}{2}(1-\frac{1}{X_0^2})} \,, \nonumber \\ 
p &\sim & \frac{1}{3} (1-4 X_0^2) \, \epsilon \,.
\end{eqnarray}
This is subleading with respect to the divergent part that scales with an exponential of $\phi$ but is still power-like divergent. 
However if we consider the regime in which the scaling variable (\ref{scalingvar}) is fixed, we have that $\phi \sim \log \tau$ and 
\be 
\epsilon \sim \tau^{-\frac{1}{3}(1-4X_0^2)} (\log \tau)^{P} \,.
\ee
We have thus a  logarithmic correction to the power-law decay of the energy density, and the result is consistent with the behaviour obtained by considering the temperature decay at the black-hole horizon and the thermodynamic relations. 
In the critical case $X_0 = -1/2$ the behaviour is purely logarithmic, and then we have a sensible behaviour only with $P < 0$. The pressure/energy ratio we obtain with this procedure is the same as for the static solution, which gives another partial confirmation of the validity of the calculation. 

\section{Conclusions and discussion}\label{Conc}

In this paper we have found the leading order effect of the presence of a trace anomaly on the approach of a system to thermal equilibrium, at least in a specific class of models described holographically by gravity coupled to a single-exponential potential, or a slight modification thereof. The main conclusion is that the deviation from conformal invariance implies a slowdown of the relaxation, encoded in a different late-time exponent that we could find analytically. 

Probably the two most pressing questions have to do with the need for a UV completion in our models, and the applicability to a real-world situation. Both questions could be answered at once by replacing the simple potential with a more complicated and phenomenologically viable one, but that would require solving the equations numerically. Alternatively, a rough estimation for the first question could be obtained by replacing the UV region of the geometry with an AdS and matching at some cutoff scale. We expect that this procedure would not influence our results for the leading behavior of the solution. Another way of getting closer to reality would be to introduce, for a generic potential, an effective parameter $X$ that would be scale-dependent; one could expect that the late-time behavior would be given by integrating the correspondent of eq. (\ref{sole}) but where the exponent $\xi$ is itself temperature-dependent, and hence time-dependent. 

It would also be obviously interesting to explore the effects of the bulk viscosity at the next order in the hydrodynamic expansions, as was done in \cite{Janik:2006ft,Heller:2007qt} for the shear viscosity, and to explore deviations from the boost-invariant assumption.

\section*{Acknowledgements}
We thank the Galileo Galilei Institute for Theoretical Physics for the 
hospitality and the INFN for partial support during the completion of 
this work. M.J. was partially supported by the
``ARISTEIA II'' Action of the  ``Operational Programme Education and Lifelong Learning'' 
and was co-funded by the European Social Fund (ESF) and National Resources.

\appendix

\section{Black brane solution the IR modified potential}
\lab{App-modsol}
The easiest way to derive the analytic solution (\ref{df})  that is valid in the region $1\ll \phi\ll\phi_h$ is to use the approach of \cite{ihqcd4}. As shown in this reference, defining the ``scalar variables'' 
$X$ and $Y$:
\begin{equation}\label{XY1}
  X(\f)\equiv \frac{1}{3}\frac{\phi'}{A'}, \qquad  Y(\phi)\equiv
  \frac{1}{4}\frac{g'}{ A'} \, .
\end{equation}
where the function $g$ is defined as $g = \log{f}$, 
 the Einstein's equations can be reduced to 
%%%%%%%%%%%%%%%%%%%%%%%%%%%%%%%%%%%%%%%%%%%%%%%%%%%%%%%%%%%%%%%%%%
\bea\label{Xeq}
\frac{dX}{d\f} &=& -\frac43~(1-X^2+Y)\le(1+\frac{3}{8}\frac{1}{X}\frac{d\log V}{d\f}\ri),\\
\frac{dY}{d\f} &=& -\frac43~(1-X^2+Y)\frac{Y}{X }. \label{Yeq} \eea
%%%%%%%%%%%%%%%%%%%%%%%%%%%%%%%%%%%%
This second order system is sufficient to determine all of the
thermodynamic properties (and dissipation) of the gravitational
theory \cite{ihqcd4}. This is a reduction of the fifth order
Einstein-scalar system to an equivalent second order system. 

It is straightforward to show that these equations combined with
the following three,
%%%%%%%%%%%%%%%%%%%%%%%%%%%%%%%%%%%%%%%%%%%%%%%%%%%%%%%%%%%%%%%%%%
\bea\label{Ap}
\frac{dA}{du} &=& -\frac{1}{\ell} e^{-\frac43\int^{\f}_{0} X(t)dt},\\
\label{fp} \frac{d\f}{du} &=&  -\frac{3}{\ell}  X(\f) e^{-\frac43\int^{\f}_{0} X(t)dt},\\
\label{gp} \frac{dg}{du} &=& -\frac{4}{\ell}~Y(\f)
e^{-\frac43\int^{\f}_{0} X(t)dt}, \eea
%%%%%%%%%%%%%%%%%%%%%%%%%%%%%%%%%%%%
solve the original Einstein equations in the domain
-wall variables defined by the Ansatz:
\begin{equation}\label{BHu}
  ds^2 = f^{-1}(r)du^2 + e^{2A(u)}\le( dx^2 + dt^2 f(r) \ri), \qquad \phi=
  \phi(u).
\end{equation}
The solution in the conformal coordinates (\ref{conf}) is found by the change of variables $du = \exp(A) dr$.

 One can also express $g$ and $A$ in terms of
the phase variables directly from the definitions (\ref{XY1}):
%%%%%%%%%%%%%%%%%%%%%%%%%%%%%%%%%%%%%%%%%%%%%%%%%%%%%%%%%%%%%%%%%%
\bea
A(\f) &=& A(\f_c) + \frac{1}{3}\int_{\f_c}^{\f} \frac{d\tilde{\f}}{X},\label{Aeq1}\\
f(\f) &=&  \exp\le(\frac43 \int_{0}^{\f} \frac{Y}{X}d\tilde{\f}\ri).\label{feq1}
\eea
%%%%%%%%%%%%%%%%%%%%%%%%%%%%%%%%%%%%%%%%%%%%%%%%%%%%%%%%%%%%%%%%%%
Here $\phi_c$ denotes some limit value. 
near the boundary where we will apply the UV matching conditions of the TG and the BH solution in the following.  
The precise form of the overall coefficient follows from inserting (\ref{Ap}), (\ref{fp}) and (\ref{gp}) in the Einstein's equations. 

Now, we apply this method to the potential  (\ref{exppotmod}). In the region $1\ll \phi\ll\phi_h$ one can solve (\ref{Xeq}) as 
\be\lab{Xsol}
X(\phi) = X_0 -\frac{3P}{8\phi} + \cO(1/\phi)\, .
\ee 
On the other hand the equation (\ref{Yeq}) can be solved for an arbitrary $X(\phi)$ as 
\be\lab{Ysolgen}
Y(\phi) = \frac{\exp\le(\int_{\phi_0}^\phi c(\phi) d\phi'\ri)}{\int_\phi^{\phi_h} d\phi' d(\phi')\exp\le(\int_{\phi_0}^{\phi'} c(\tilde{\phi}) d\tilde{\phi}\ri) }\, ,
\ee
where we defined 
\be\lab{cd}
c(\phi) = \frac{4(X(\phi)^2-1)}{3X(\phi)}, \qquad d(\phi) = -\frac{4}{3X(\phi)}\, .
\ee
Substituting (\ref{Xsol}) in (\ref{cd}) and (\ref{Ysolgen}) and expanding for large $\phi$, one finds 
\be\lab{Ysol1}
Y(\phi) = \frac{1-X_0^2}{ \exp\le[a(\phi-\phi_h)\ri] (\phi/\phi_h)^b}\le(1 + \cO(1/\phi) \ri)\, ,
\ee
where $a = 4(1-X_0^2)/(3X_0)$ and $b = P(1 + 1/X_0^2)/2$. One then uses (\ref{Aeq1}) and (\ref{feq1}) and finds the solution (\ref{df}). 
One can  finally obtain the solution in the domain-wall coordinates by solving (\ref{fp}) in the large $\phi$ limit as 
\be\lab{phiu}
u-u_0= \frac{3}{4X_0 \tilde{c}}\exp\le(4X_0\phi/3\ri)\phi^{-P/2}\le(1+ \cO(1/\phi)\ri)\, ,
\ee
where we defined
\be\lab{defc}
\tilde{c} = -\frac{3X_0}{\ell} \exp\le(\frac43 X_0\phi_0\ri)\phi_0^{-P/2}\, . 
\ee
Finally one can invert this equation in the large $\phi$ limit to obtain (\ref{phiumod}).

\section{Free energy from on-shell gravity action} 
\lab{App-FE}

In this appendix we evaluate the difference of the on-shell actions 
between the black-brane and the thermal gas and   we prove that the analytic 
solutions describe above do not demonstrate
a Hawking-Page transition. The action is given by (\ref{action}). One finds that 
the trace of the
intrinsic curvature is given by,
\begin{equation}\label{intcuru}
    K = \frac{\sqrt{f}}{2}(8A'+f'/f)
\end{equation}
in the domain-wall coordinate system. Thus, the boundary
contribution to the action becomes,
\begin{equation}\label{sboun}
    S_{bnd} = M^3 V_3 \b \le\{ e^{g+4A}(8A'+f'/f)\ri\}_{u_b},
\end{equation}
where $u_b$ denotes the regulated boundary of the geometry
infinitesimally close to $-\infty$.

 The bulk contribution to the action, evaluated on the
solution can be simplified as,
\begin{eqnarray}
  S_{bulk} &=& 2 M^3 V_3 \b \int_{u_b}^{u_s} du \frac{d}{du}\le(f e^{4A} A'\ri) 
\nonumber \\
  {} &=& 2 M^3 V_3 \b \le\{f(u_s)e^{4A(u_s)}A'(u_s) -
  f(u_b)e^{4A(u_b)}A'(u_b)\ri\}\label{sbulk}\,.
\end{eqnarray}
Here $u_s$ denotes $u_0$ or $u_h$ depending on which appears first.
Thus, for the black-hole solution $u_s=u_h$, whereas for the thermal
gas $u_s = u_0$.

The first term in (\ref{sbulk}) deserves attention. Clearly it
vanishes for the black-hole, as $f(u_h)=0$ by definition. However,
it is not a priori clear that it also vanishes for the thermal gas.
A straightforward computation using (\ref{Asol}),(\ref{dilsoll})
and,
\begin{equation}\label{Apsol}
    A' = -\frac{1}{\ell}\l^{-\frac{4X}{3}}
\end{equation}
shows that it indeed vanishes for our physically interesting case
$X^2<1$. Therefore, one obtains the following total expression for
the action from (\ref{sboun}) and (\ref{sbulk}) by dropping the
first term in (\ref{sbulk}):
\begin{equation}\label{stotl}
    S = M^3V_3\b e^{g(u_b)+4A(u_b)}\le(6A'(u_b)+ g'(u_b)\ri).
\end{equation}

In order to compare the energies of the black-hole and the thermal
gas geometries, we fix the UV asymptotics of the thermal gas
geometry by requiring the same circumference for the Euclidean time
at $u_b$:
\begin{equation}\label{tfix}
    \bar{\b} = \b \sqrt{f(u_b)}\,.
\end{equation}

Now, it is straightforward to compute the energy of the geometries.
For the black-hole (\ref{BHmet}), one finds:
\begin{equation}\label{sBH}
    S_{BH} = 2M^3 V_3\le(\frac{\b}{\ell}\ri) e^{4A_0} 
\le(C_2(1+2X^2)-3\l_b^{\a}\ri)\,.
\end{equation}
Here $\l_b$ is the value of the dilaton on the regulated boundary
$u_b$ and $\a = 4(1-X^2)/(3X)$. As $\a<0$ and $\l\to 0$ near the boundary, it is a divergent
piece that should be regulated.

For the thermal gas one finds, using (\ref{tfix}),
\begin{equation}\label{sTG}
    S_{TG} = 3 M^3 V_3\le(\frac{\b}{\ell}\ri) e^{4A_0} \le(C_2 - 
2 \l_b^{\a}\ri)\,.
\end{equation}
We note that the divergent terms in (\ref{sBH}) and (\ref{sTG})
cancel in the difference and one finds,
\begin{equation}\label{sdif}
    S_{BH} - S_{TG} = - M^3 V_3\le(\frac{\b}{\ell}\ri) e^{4A_0} 
C_2\le(1 -4 X^2\ri)\,.
\end{equation}
We note from (\ref{temp}) that the temperature is given by,
\begin{equation}\label{tempfin}
    e^{A_0} = \frac{\pi T \ell}{1-X^2}C_2^{-\frac{\frac14-X^2}{1-X^2}}\,.
\end{equation}
By using this relation, we see that the difference~\eqref{sdif} indeed agrees with the free energy found by integrating the black hole entropy in~\eqref{free}.

\section{General solution to the Einstein equations}\label{App-Gensol}

By inserting\footnote{It turns out to be useful to change variables from $z$ to $v$ before deriving the equations.} the Ansatz of~\eqref{dabc} and~\eqref{laansatz} 
in the Einstein equations~\eqref{Einsteqs}, and after choosing the gauge $d=0$, one obtains at leading order in $1/\tau$ the system
\begin{align} \label{nleq1}
0=&\,\frac{v \left(1-4 X^2\right) \left(a'(v) b'(v)+2 a'(v) c'(v)+2 b'(v) c'(v)+c'(v)^2\right)}{16 } \\ \nn
&-\frac{3 \left(a'(v)+b'(v)+2 c'(v)\right)}{8 }
+ X\lambda_1'(v)
-\frac{v \left(1-4 X^2\right) \lambda_1'(v)^2}{6}\\ \nn
&+\frac{3 \left(1-X^2\right) \left(1-e^{-8 X \lambda_1(v)/3}\right)}{2 v  \left(1-4 X^2\right)}  \\[4mm]
0=&-\frac{1}{2} a'(v) \left(b'(v)+2 c'(v)\right)+\frac{3 a'(v)}{v \left(1-4 X^2\right)} \\ \nn
&+b''(v)+\frac{1}{2} b'(v)^2+2 c''(v)+c'(v)^2+\frac{8}{3} \l_1'(v)^2
-\frac{16 X \l_1'(v)}{v-4 v X^2}  \\[4mm] 
0=&\,a''(v)-\frac{1}{2} b'(v) \left(a'(v)+2 c'(v)\right)+\frac{1}{2} a'(v)^2+\frac{3 b'(v)}{v-4 v X^2} \\ \nn
&+2 c''(v)+c'(v)^2+\frac{8}{3} \l_1'(v)^2-\frac{16 X \l_1'(v)}{v-4 v X^2} \\[4mm]
0=&\,\frac{1}{2} a'(v) \left(b'(v)-c'(v)\right)+b''(v)-\frac{3 \left(b'(v)-c'(v)\right)}{v-4 v X^2} \\ \nn
&+\frac{1}{2} b'(v) c'(v)+\frac{1}{2} b'(v)^2-c''(v)-c'(v)^2  \\[4mm]
\label{nleq5}
0=&\left(3 s-4+16 X^2\right) a'(v)-(s-4) \left(1-4 X^2\right) b'(v) \\ \nn
&-2 s \left(1-4 X^2\right) c'(v)-8 s X \l_1'(v)  \, .
\end{align}
The number of equations exceeds the number of variables by one, but the system is not overconstrained: any of the second order equations can be derived from the other equations. 
Notice that this system approaches smoothly that found in~\cite{Janik:2005zt} as $X \to 0$ (so that the CR solution becomes the AdS$_5$ solution).

\subsection{UV behavior}

Let us first discuss the behavior of the solutions to~\eqref{nleq1}--\eqref{nleq5}  near the UV boundary. %We start from the full system, given in~\eqref{nleq1}--\eqref{nleq5} in the main text. 
Inserting here an Ansatz where all fields have the behavior $\sim \mathrm{const.}\times v^\Delta$, and requiring that a solution exists at small $v$, we recover the characteristic equation
\be
 \Delta ^3 \left(-\Delta +4 \Delta  X^2-4 X^2+4\right)^2 \left(-\Delta +4 \Delta  X^2+4 X^2\right) = 0 \, .
\ee
Notice that this is a sixth order equation, reflecting the number of integration constant in a general solution to~\eqref{nleq1}--\eqref{nleq5}. There is the triple root at $\Delta=0$, but this simply reflects the fact that the equations are trivially solved by constant functions $a$, $b$ and $c$ as only their derivatives appear.
The nontrivial solutions are
\begin{enumerate}
 \item A single root at $\Delta = \frac{4X^2}{1-4X^2}$. In terms of the basis of the functions $A$, $m$, and $n$ defined in~\eqref{basis1}--\eqref{basis4}, 
 only the function $A$ is nonzero (at leading order for small $v$) for this solution. Notice that $\Delta \to 0$ in the conformal limit $X \to 0$. 
 \item A double root at $\Delta = \frac{4(1-X^2)}{1-4X^2} \equiv \xi$. 
 In the conformal limit $\Delta \to 4$, and this solution is therefore identified with turning on a finite energy-momentum tensor.  
 For this solution all functions $A$, $m$, and $n$ are nontrivial and the solution is parametrized in terms of two integration constants:
 \bea \label{UVexps}
  m &=& m_c v^\xi + \mathcal{O}\le(v^{2\xi}\ri)\,,\quad n = n_c v^\xi + \mathcal{O}\le(v^{2\xi}\ri)\,,\nonumber\\
  A &=& \frac{4 X \left[X(1-4X^2) m_c+ (1-X^2)n_c\right]}{3 \left(1-2 X^2\right)}v^\xi + \mathcal{O}\le(v^{2\xi}\ri)\,.
 \eea
\end{enumerate}

\subsection{Analytic solution}

Let us then discuss general analytic solutions to the system~\eqref{nleq1}--\eqref{nleq5}. Recall from the main text that $A$ and $n$ could be eliminated in terms of $m$: 
\be \label{Ansolapp}
 A(w) = \frac{\xi}{2}(w-w_0) - \frac{1}{2} \log \le|\frac{2 \S m'(w)}{\xi}\ri| \, , \qquad n(w) = \k\, m(w) + n_0\,.
\ee
where $\xi=4(1-X^2)/(1-4X^2)$, and $\S$ is given in~\eqref{Sdef}. The additional parameters $w_0$ and $n_0$ are constants of integration. %We choose to fix them such that for the solution with $\Delta = 4(1-X^2)/(1-4X^2)$, all functions tend to zero as the UV boundary is approached. 
The remaining equations take the form
\begin{align}
\label{peqapp}
&\phantom{+} \frac{8 \left(1-X^2\right) X^4}{\left(1-4 X^2\right)^2}+4 \frac{K  X^2}{1-4X^2} p(w)-\frac{\S^2-K ^2}{2 \left(1-X^2\right)}p(w)^2
\\\nn
&+K p'(w)+ \frac{2 X^2-4 X^4}{1-4 X^2 }\frac{p'(w)}{p(w)}-\frac{1+X^2 }{2 }\frac{p'(w)^2}{p(w)^2}
=-\frac{p''(w)}{2 p(w)}\,,\\
&&\nonumber\\
\label{constreq}
&\phantom{+} \frac{8 \exp \left[-2 K m(w)+2 X^2 \log \left|2 \S m'(w)/\xi\right|-\frac{8}{3} n_0 \left(1-X^2\right) X- 2 \xi X^2 (w-w_0)\right]}{\left(1-4 X^2\right)^2}\nn\\
=&\phantom{+} \frac{8 X^4}{\left(1-4 X^2\right)^2}+\frac{\left(K^2-\S^2\right) m'(w)^2}{2 \left(1-X^2\right)^2}+\frac{4 K X^2\, m'(w)}{(1-X^2)(1-4X^2)}+ \frac{K}{1-X^2}m''(w)\\\nn
&+\frac{4 X^2}{\left(1-4 X^2\right)}\frac{m''(w)}{m'(w)}+\frac{m''(w)^2}{2\, m'(w)^2}\,,
\end{align}
where $p(w)=m'(w)$ and the constants $\S$ and $K$ were defined in~\eqref{Sdef}. It is straightforward to check that these two equations are equivalent up to the choice of integration constants in the second equation.
Further defining 
\be \label{gammadef}
 \gamma(w) = -\frac{K}{X^2}+\frac{p'(w)}{p(w)^2}+\frac{4 X^2}{1-4 X^2}\frac{1}{p(w)}
\ee
and by using $m$ as the variable,~\eqref{peqapp} may be written as
\be \label{gammaeq}
\gamma'(m) +\frac{2 K \gamma (m)}{X^2}+\left(1-X^2\right) \gamma (m)^2 + \frac{K^2-\S^2 X^4}{X^4(1-X^2)}= 0\,.
\ee
The equations~\eqref{gammaeq} and~\eqref{gammadef} can be easily solved:
\bea
 \gamma(m) &=& \frac{\S X^2\, \coth (\S (m-m_0))-K}{X^2 \left(1-X^2\right)}\\
\label{psol}
 p(m) &=& C_p e^{-\frac{K m}{1-X^2}}\le|\sinh (\S (m-m_0))\ri|^\frac{1}{1-X^2}\\\nn
  &&+\frac{\xi}{2\S} \left(e^{2 \S (m-m_0)}-1\right) \, _2F_1\left(1,\frac{ \S(1-2 X^2)+K}{2 \S(1- X^2)};\frac{1-2 X^2}{1-X^2};1-e^{2 \S(m-m_0)}\right)
\eea
Another branch of (real valued) solutions is obtained from here by shifting $m_0 \to m_0 + \pi i/(2\S)$ 
and by choosing a suitable branch for the hypergeometric function, but as it turns out, these solutions do not admit regular UV boundaries. 
Therefore we will not discuss them further.

Let us then comment on the integration constants. As pointed out above there are three trivial constants because the equations depended on $a$, $b$ and $c$ only through their derivatives. 
Two of these constants are identified as $n_0$ and $m_0$, and the third one was already fixed implicitly by the change of basis in~\eqref{basis1}--\eqref{basis4}, 
where an additional constant could be added in such a way that~\eqref{nleq5} is still automatically satisfied.
A convenient choice is to set $m_0=0=n_0$: this will ensure that all functions vanish at the boundary. 
Further the invariance of the system~\eqref{nleq1}--\eqref{nleq5} in rescalings $v \to C v$ can be used to set $w_0=0$.
The remaining nontrivial integration constants are identified as $\kappa$ (or equivalently $K$), and $C_p$. 

Remarkably,~\eqref{psol} can be further integrated to give $w$ in terms of $m$. This is perhaps most easily done by substituting the result for $p$ in the constraint equation~\eqref{constreq}, using the relation
\be
a z (1-z) \, _2F_1(2,a+1;b+1;z)+b (-a z+b-1) \, _2F_1(1,a;b;z)+b(1-b) =0
\ee
for the hypergeometric functions to simplify the expression, and solving for $w$. The result reads
\bea \label{wsolgeneral}
 w &=& -\frac{1-4X^2}{4X^2} \log \Bigg[\, _2F_1\left(1,\frac{ \S(1-2 X^2)+K}{2 \S(1- X^2)};\frac{1-2 X^2}{1-X^2};1-e^{2 \S m}\right)\\\nn
 &&+C_p\frac{2 \S  e^{-\frac{K m}{1-X^2}} \le|\sinh (\S m)\ri|^\frac{1}{1-X^2}}{\xi \left|e^{2 m \S}-1\right|}
 \Bigg]+\frac{1}{\xi}\log \left|e^{2 m \S}-1\right|-\frac{  (K+\S)}{\xi X^2}m\,.
\eea
By inserting the solution in the expression for $A$ in~\eqref{Ansolapp} and expanding at the UV (where $w \to \infty$ and $m \to 0$), we see that a nonzero constant $C_p$ corresponds to turning on the mode 1.
of the previous section with $\Delta=4X^2/(1-4X^2)$. Since we only want to turn on a nonzero energy-momentum tensor, we will set $C_p=0$. For this choice we recover the expansions in~\eqref{UVexps} with
\be
 m_c = \frac{e^{-\xi w_0}}{2 \S} = \frac{1}{2 \S}\ , \qquad n_c = \kappa m_c \,.
\ee

After setting $C_p=0$, the solution~\eqref{wsolgeneral} may become singular at the zeroes of the hypergeometric function. However, as seen in the main text, the hypergeometric function cancels when the leading order metric is expressed by using $m$ as the bulk variable instead of $w$ or $v$. Therefore the zeroes of the hypergeometric function are coordinate singularities, 
whereas curvature singularities may arise at infinite $m$.

The resulting solution has two branches, one where $m$ runs from zero at the boundary to $+\infty$ and the other where $m$ runs from zero to $-\infty$. 
The branch with positive values of $m$ was discussed in the main text. The branch with negative $m$ can be analyzed similarly, in this case the absence of curvature singularity cannot be avoided.
More precisely, following the procedure of the main text, the absence of the singularity at $m=-\infty$ would imply that 
\be
 s=4\,,\qquad \kappa = -4 X\,
\ee
instead of~\eqref{IRregconds}. The value of $s=4$ conflicts, however, with our initial assumptions: the expansion in $1/\t$ would break down for this value.


\begin{thebibliography}{99}

 %\cite{Janik:2005zt}
\bibitem{Janik:2005zt} 
  R.~A.~Janik and R.~B.~Peschanski,
  %``Asymptotic perfect fluid dynamics as a consequence of Ads/CFT,''
  Phys.\ Rev.\ D {\bf 73}, 045013 (2006)
  [hep-th/0512162].
  %%CITATION = HEP-TH/0512162;%%
% 
\bibitem{Bjorken:1982qr}
  J.~D.~Bjorken,
  %``Highly Relativistic Nucleus-Nucleus Collisions: The Central Rapidity Region,''
  Phys.\ Rev.\ D {\bf 27} (1983) 140.
  %%CITATION = PHRVA,D27,140;%%

%\cite{Romatschke:2009im}
\bibitem{Romatschke:2009im}
  P.~Romatschke,
  %``New Developments in Relativistic Viscous Hydrodynamics,''
  Int.\ J.\ Mod.\ Phys.\ E {\bf 19} (2010) 1
  [arXiv:0902.3663 [hep-ph]].
  %%CITATION = ARXIV:0902.3663;%%
%\cite{Chamblin:1999ya}

%\cite{Policastro:2001yc}
\bibitem{Policastro:2001yc}
  G.~Policastro, D.~T.~Son and A.~O.~Starinets,
  %``The Shear viscosity of strongly coupled N=4 supersymmetric Yang-Mills plasma,''
  Phys.\ Rev.\ Lett.\  {\bf 87} (2001) 081601
  [hep-th/0104066].
  %%CITATION = HEP-TH/0104066;%%
  %978 citations counted in INSPIRE as of 29 Jul 2015

%\cite{Kovtun:2004de}
\bibitem{Kovtun:2004de}
  P.~Kovtun, D.~T.~Son and A.~O.~Starinets,
  %``Viscosity in strongly interacting quantum field theories from black hole physics,''
  Phys.\ Rev.\ Lett.\  {\bf 94} (2005) 111601
  [hep-th/0405231].
  %%CITATION = HEP-TH/0405231;%%
  %1486 citations counted in INSPIRE as of 29 Jul 2015

%\cite{Buchel:2003tz}
\bibitem{Buchel:2003tz}
  A.~Buchel and J.~T.~Liu,
  %``Universality of the shear viscosity in supergravity,''
  Phys.\ Rev.\ Lett.\  {\bf 93} (2004) 090602
  [hep-th/0311175].
  %%CITATION = HEP-TH/0311175;%%
  %356 citations counted in INSPIRE as of 29 Jul 2015
  
%\cite{Cremonini:2011iq}
\bibitem{Cremonini:2011iq}
  S.~Cremonini,
  %``The Shear Viscosity to Entropy Ratio: A Status Report,''
  Mod.\ Phys.\ Lett.\ B {\bf 25} (2011) 1867
  [arXiv:1108.0677 [hep-th]].
  %%CITATION = ARXIV:1108.0677;%%
  %40 citations counted in INSPIRE as of 29 Jul 2015  

%\cite{Gursoy:2007cb}
\bibitem{ihqcd1}
  U.~Gursoy and E.~Kiritsis,
  %``Exploring improved holographic theories for QCD: Part I,''
  JHEP {\bf 0802} (2008) 032
  [arXiv:0707.1324 [hep-th]].
  %%CITATION = ARXIV:0707.1324;%%
  %212 citations counted in INSPIRE as of 25 May 2015
  
%\cite{Gursoy:2007er}
\bibitem{ihqcd2}
  U.~Gursoy, E.~Kiritsis and F.~Nitti,
  %``Exploring improved holographic theories for QCD: Part II,''
  JHEP {\bf 0802} (2008) 019
  [arXiv:0707.1349 [hep-th]].
  %%CITATION = ARXIV:0707.1349;%%
  %223 citations counted in INSPIRE as of 25 May 2015


\bibitem{ChamblinReall}
  H.~A.~Chamblin and H.~S.~Reall,
  %``Dynamic dilatonic domain walls,''
  Nucl.\ Phys.\  B {\bf 562} (1999) 133
  [hep-th/9903225].
  %%CITATION = NUPHA,B562,133;%%
  
 
  
  %\cite{Janik:2006ft}
\bibitem{Janik:2006ft}
  R.~A.~Janik,
  %``Viscous plasma evolution from gravity using AdS/CFT,''
  Phys.\ Rev.\ Lett.\  {\bf 98} (2007) 022302
  [hep-th/0610144].
  %%CITATION = HEP-TH/0610144;%%
  
%\cite{Bak:2006dn}
\bibitem{Bak:2006dn} 
  D.~Bak and R.~A.~Janik,
  %``From static to evolving geometries: R-charged hydrodynamics from supergravity,''
  Phys.\ Lett.\ B {\bf 645}, 303 (2007)
  [hep-th/0611304].
  %%CITATION = HEP-TH/0611304;%%
  
  %\cite{Gouteraux:2011qh}
\bibitem{Gouteraux:2011qh}
  B.~Gouteraux, J.~Smolic, M.~Smolic, K.~Skenderis and M.~Taylor,
  %``Holography for Einstein-Maxwell-dilaton theories from generalized dimensional reduction,''
  JHEP {\bf 1201} (2012) 089
  [arXiv:1110.2320 [hep-th]].


%\cite{Gursoy:2008bu}
\bibitem{ihqcd3}
  U.~Gursoy, E.~Kiritsis, L.~Mazzanti and F.~Nitti,
  %``Deconfinement and Gluon Plasma Dynamics in Improved Holographic QCD,''
  Phys.\ Rev.\ Lett.\  {\bf 101} (2008) 181601
  [arXiv:0804.0899 [hep-th]].
  %%CITATION = ARXIV:0804.0899;%%
  %120 citations counted in INSPIRE as of 25 May 2015

%\cite{Gursoy:2008za}
\bibitem{ihqcd4}
  U.~Gursoy, E.~Kiritsis, L.~Mazzanti and F.~Nitti,
  %``Holography and Thermodynamics of 5D Dilaton-gravity,''
  JHEP {\bf 0905} (2009) 033
  [arXiv:0812.0792 [hep-th]].
  %%CITATION = ARXIV:0812.0792;%%
  %134 citations counted in INSPIRE as of 25 May 2015

%\cite{Gursoy:2009jd}
\bibitem{ihqcd5}
  U.~Gursoy, E.~Kiritsis, L.~Mazzanti and F.~Nitti,
  %``Improved Holographic Yang-Mills at Finite Temperature: Comparison with Data,''
  Nucl.\ Phys.\ B {\bf 820} (2009) 148
  [arXiv:0903.2859 [hep-th]].
  %%CITATION = ARXIV:0903.2859;%%
  %85 citations counted in INSPIRE as of 25 May 2015
  
    
%\cite{Papadimitriou:2011qb}
\bibitem{Papadimitriou:2011qb}
  I.~Papadimitriou,
  %``Holographic Renormalization of general dilaton-axion gravity,''
  JHEP {\bf 1108} (2011) 119
  [arXiv:1106.4826 [hep-th]].

\bibitem{papadim-private}
 I.~Papadimitriou, private communication
 
 %\cite{Ishii:2015gia}
\bibitem{Ishii:2015gia}
  T.~Ishii, E.~Kiritsis and C.~Rosen,
  %``Thermalization in a Holographic Confining Gauge Theory,''
  arXiv:1503.07766 [hep-th].
  %%CITATION = ARXIV:1503.07766;%%
  
  %\cite{Buchel:2015ofa}
\bibitem{Buchel:2015ofa}
  A.~Buchel and A.~Day,
  %``Universal relaxation in quark-gluon plasma at strong coupling,''
  arXiv:1505.05012 [hep-th].
  %%CITATION = ARXIV:1505.05012;%%
  
  %\cite{Janik:2015waa}
\bibitem{Janik:2015waa}
  R.~A.~Janik, G.~Plewa, H.~Soltanpanahi and M.~Spalinski,
  %``Linearized nonequilibrium dynamics in nonconformal plasma,''
  Phys.\ Rev.\ D {\bf 91} (2015) 12,  126013
  [arXiv:1503.07149 [hep-th]].
  %%CITATION = ARXIV:1503.07149;%%

  
%\cite{Mas:2007ng}
\bibitem{TarrioMas}
  J.~Mas and J.~Tarrio,
  %``Hydrodynamics from the Dp-brane,''
  JHEP {\bf 0705} (2007) 036
  [hep-th/0703093].
  %%CITATION = HEP-TH/0703093;%%
  %44 citations counted in INSPIRE as of 14 juil. 2015

%\cite{Eling:2011ms}
\bibitem{ElingOz}
  C.~Eling and Y.~Oz,
  %``A Novel Formula for Bulk Viscosity from the Null Horizon Focusing Equation,''
  JHEP {\bf 1106} (2011) 007
  [arXiv:1103.1657 [hep-th]].
  %%CITATION = ARXIV:1103.1657;%%
  %20 citations counted in INSPIRE as of 14 Jul 2015
  
  %\cite{Heller:2007qt}
\bibitem{Heller:2007qt}
  M.~P.~Heller and R.~A.~Janik,
  %``Viscous hydrodynamics relaxation time from AdS/CFT,''
  Phys.\ Rev.\ D {\bf 76} (2007) 025027
  [hep-th/0703243].
  %%CITATION = HEP-TH/0703243;%%

\end{thebibliography}
\end{document}